\documentclass[3p,times,twocolumn]{elsarticle}

\usepackage{times}
\usepackage{lscape}
\usepackage{ecrc}

\volume{00}

\firstpage{1}

\journalname{Journal of High Energy Astrophysics}
\journal{Journal of High Energy Astrophysics}

\runauth{G. Ghisellini}

\jid{nuphbp}

 \jnltitlelogo{\Large Journal of High Energy Astrophysics}

\usepackage{amssymb}

\usepackage[figuresright]{rotating}

\def\gsim{ \lower .75ex \hbox{$\sim$} \llap{\raise .27ex \hbox{$>$}} }
\def\lsim{ \lower .75ex\hbox{$\sim$} \llap{\raise .27ex \hbox{$<$}} }

\def\beq{\begin{equation}}
\def\eeq{\end{equation}}

\begin{document}

\begin{frontmatter}



\dochead{}

\title{Swift for blazars}


\author{Gabriele Ghisellini}

\address{INAF -- Osservatorio Astronomico di Brera, Via Bianchi 46, Merate, Italy}

\begin{abstract}

I will review recent advances in the field of blazars, 
highlighting the contribution of {\it Swift}.
Together with other operating satellites (most notably {\it Fermi},
but also {\it AGILE}, {\it WISE}, {\it Planck}) and ground based facilities
such as Cherenkov telescopes, {\it Swift} was (and is) crucial for improving our
understanding of blazars.
The main advances in the blazar field made possible by {\it Swift} includes
the opening of the time domain investigation, since there are several sources 
with hundreds of simultaneous optical, UV and X--ray data taken at different times;
the possibility to measure the black hole mass in very powerful blazars,
that show clear signs of accretion disk emission;
the possibility to classify blazar candidates, through X--ray observations; 
the finding of the most powerful and distant blazars, emitting strongly in 
the hard X--ray band accessible to {\it Swift}/BAT.
All these improvements had and have a great impact on our understanding
on how relativistic jets are formed and emit, on their power, and on how the heavy black holes
in these systems first formed and grew.
\end{abstract}
\end{frontmatter}
~

\section{Introduction}

After the launch of the {\it Compton Gamma Ray Observatory, CGRO}, we
discovered that blazars (i.e. quasars with jets pointing nearly at us) 
were the most important class of persistent $\gamma$--ray emitters.
This was somewhat unexpected, despite the fact that the previous $\gamma$--ray
satellite {\it COSB} already detected 3C 273 as a $\gamma$--ray source
\cite{swanenburg78}, \cite{bignami79}.
It was not completely clear that the production of $\gamma$--rays was associated
to relativistic jets, even if all the necessary ingredients were known 
since the early seventies: superluminal motion and the presence of
relativistic electrons in the source, producing synchrotron and self--Compton radiation
(the external Compton idea was yet to come).

The discovery of the the strong $\gamma$--ray emission of 3C 279 in 1991 by {\it CGRO}
was soon followed by the realisation that blazars are $\gamma$--ray emitters as a class,
and this triggered a frantic phase of theoretical developments \cite{maraschi92}, 
\cite{dermer93}, \cite{sikora94}, \cite{ghisellini96}, \cite{bloom96}.
At the same time, the (few)

~

\noindent
multi--wavelengths campaigns showed coordinated
variability of the flux at different frequencies, and this made the jet 
paradigm to shift from a multi--zone jet, producing the highest
frequencies ($\gamma$--rays) in the innermost regions and the 
IR--optical further out \cite{marscher80}, \cite{konigl81}, \cite{ghisellini85}, 
to the much simpler ``one--zone" jet
in which most of the emission was produced in a single region,
i.e. the same electron population producing the synchrotron
was also responsible for the high energy flux (but not the radio,
due to the synchrotron self--absorption).
This required a strong effort, especially on the observational side,
because it was not easy to organise multiwavelength campaigns
joining space and ground observatories.

When {\it Swift} was launched what was a dream became routine:
simultaneous optical, UV, and X--rays observations became easily accessible and
flexible planning allowed to use Target of Opportunity 
observations to follow extraordinary events.
Then, when {\it Fermi} joined in, we could have a really complete view of the
behaviour of blazars, and not only of the 3 or 4 brightest ones, but of
hundreds of them.

The {\it Swift} and {\it Fermi} satellite, together with ground based facility
like the Cherenkov telescopes, made a quantum jump
in our knowledge if the physics of blazars, and led the way to use them
not only to understand the high energy physical processes that characterise
their emission, but to use blazars as a probe of the far Universe.

What follows is a partial view of the recent advances in blazar science
allowed by {\it Swift}.

\section{Multi--wavelength campaigns}

\begin{figure}
\vskip -1.2 cm
\hskip -0.1 cm
\includegraphics[width=8.5cm,clip]{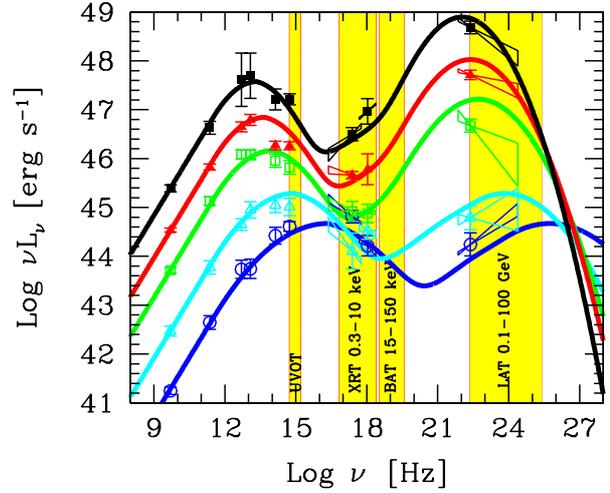}
\vskip -1. cm
\caption{
The ``blazar sequence": blazars have SEDs that changes according
to the bolometric observed jet luminosity. 
Low powerful lineless BL Lacs are ``blue":
their synchrotron and Compton hump peaks at high frequencies, and
the corresponding luminosities are about equal.
Powerful flat spectrum radio quasars (with broad emission lines) are redder,
and the Compton hump dominates.
This has been explained as due to radiative cooling:
electrons in more powerful sources suffer more severe losses, and this
limits their typical energies to values smaller than the one in low powerful
BL Lacs, in which the cooling is less severe \cite{ghisellini98}.
In this respect, the presence or absence of the broad emission lines
can play a crucial role, since they can largely enhance the inverse Compton emission
and the corresponding radiative cooling. 
The indicated yellow vertical stripes correspond to the observing bands
of {\it Swift} and {\it Fermi}/LAT. 
Adapted from \cite{fossati98}, \cite{donato01}.
}
\label{gfos}        
\end{figure}

Both planned observations together
with other instruments and target of opportunity (ToO) observations 
(performed after even a very short notice) have secured the optical--UV and X--ray
observations of hundreds of blazars. 
Fig. \ref{gfos} shows  the observed spectral energy distribution (SED) of 
blazars together with the observing band of {\it Swift} and {\it Fermi}/LAT.
{\it Swift}/UVOT and XRT cover the peak of the synchrotron emission
of low power line--less BL Lacs, the best candidate to be TeV emitters,
while {\it Swift}/BAT can be more effective in observing and even discover
powerful flat spectrum radio quasars (FSRQ) with broad emission lines
at high redshift. 
For this class of objects, {\it Swift}/UVOT can observe the thermal emission produced
by their accretion disk.

\begin{figure}
\vskip -0.2 cm
\hskip 1 cm
\includegraphics[width=6.3cm,clip]{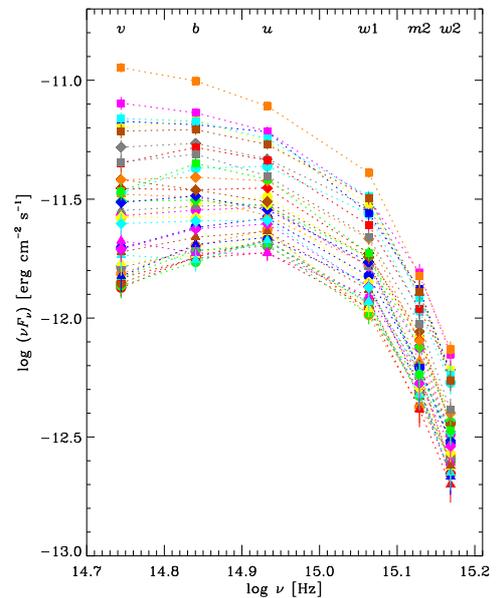}
\caption{
The {\it Swift}/UVOT SED of the blazar B3 1633+382 (alias 4C38.41),
at different epochs. 
The flux is the sum of the steep tail of the synchrotron jet emission and the
thermal component produced by the accretion disk. 
From \cite{raiteri12}.
}
\label{raiteri}       
\end{figure}

This thermal component may be elusive for intermediate redshift and intermediate power
FSRQs, because in these objects the beamed non--thermal flux can hide the thermal continuum.
One example 3C 454.3 \cite{vercellone09}, \cite{bonnoli11}, \cite{raiteri08}, whose
thermal emission was revealed through the optical and UV monitoring involving {\it Swift}/UVOT \cite{raiteri11}.
Another example is B3 1633+382 ($z=1.814$), discovered as a $\gamma$--ray source by CGRO 
and well monitored by {\it Swift}, {\it Fermi} and {\it AGILE} \cite{raiteri12}.
As Fig. \ref{raiteri} shows, {\it Swift} was instrumental to reveal the contribution
of the accretion disk.

\begin{figure}
\vskip -1 cm
\hskip -0.5 cm
\includegraphics[width=8.4cm,clip]{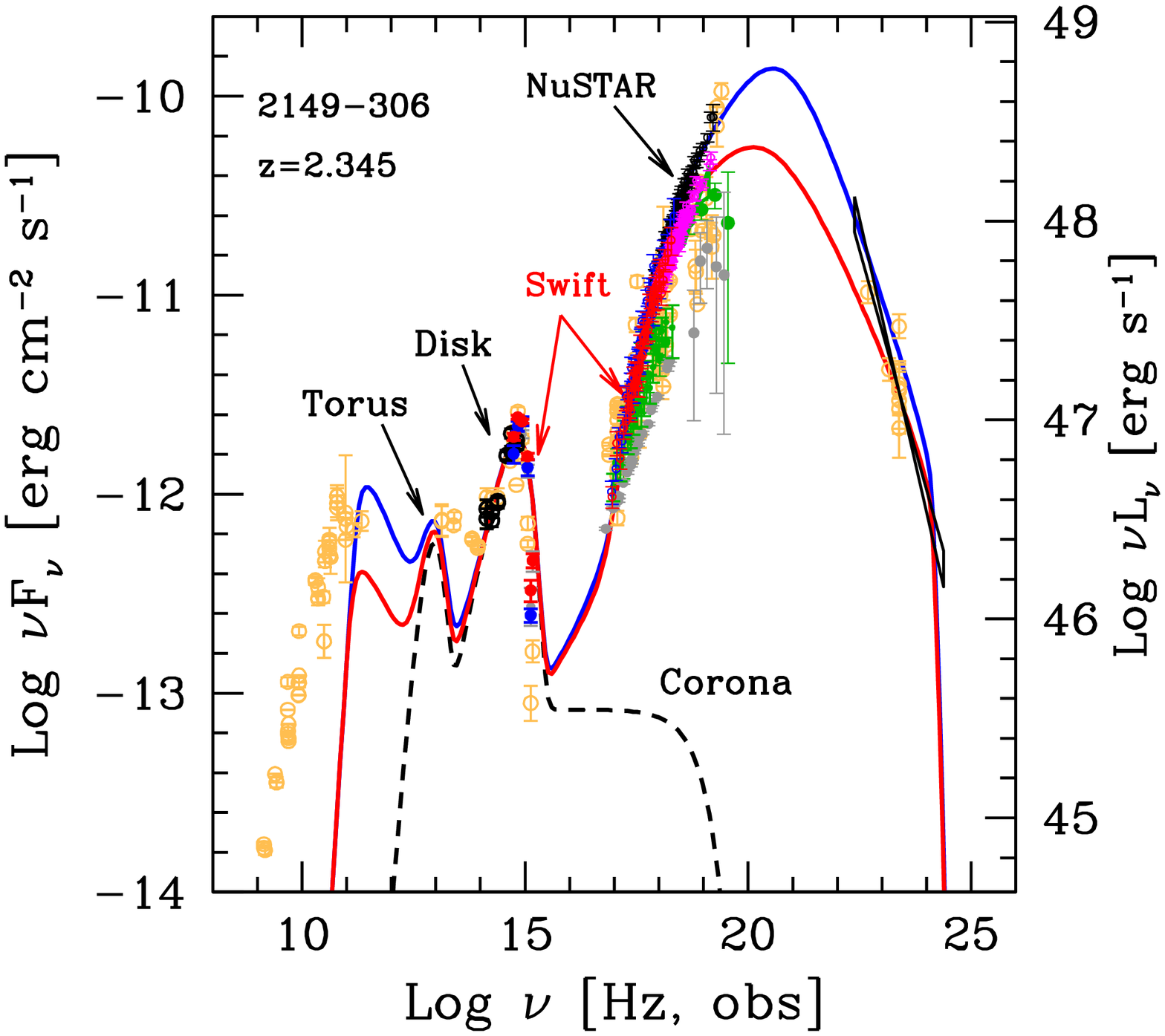} 
\vskip -0.5 cm
\hskip -0.2 cm
\includegraphics[width=8.6cm,clip]{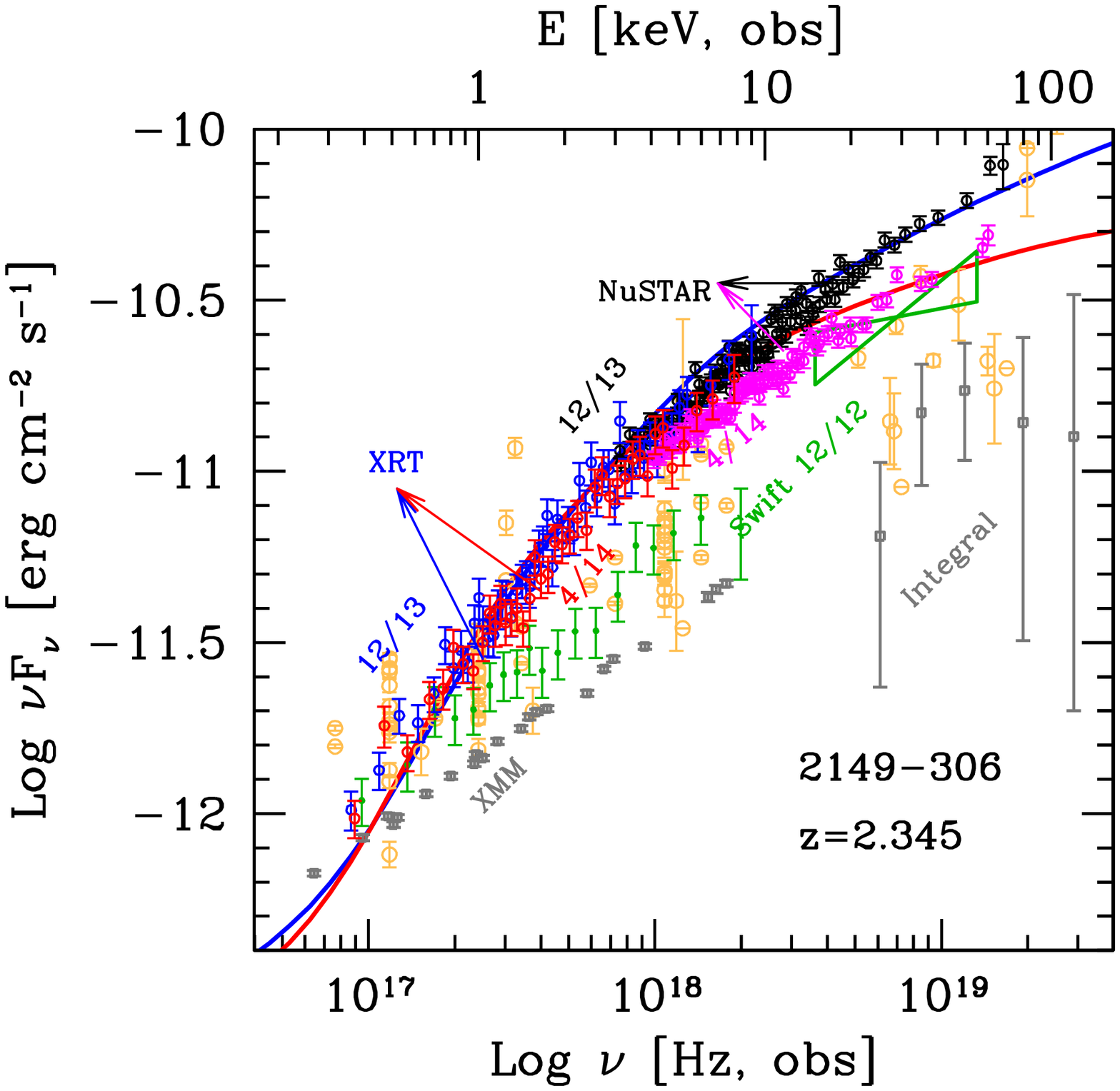}
\vskip -0.7 cm
\caption{
{\it Swift}/XRT and {\it NuSTAR}, observing simultaneously,
 cover the 0.3--70 keV band.
{\bf The top panel} shows the overall SED of PKS 2149--306, a blazar at $z=2.345$:
the optical--UV bump is due to the accretion disk, while the extremely strong and hard X--ray emission
is due to the beamed jet emission, and it is interpreted as inverse Compton scattering off the 
photons produced mainly by the infrared torus (responsible for the hump at $\sim 10^{13}$ Hz). 
{\bf The bottom panel} is a zoom on the X--ray portion of the SED, showing how the spectrum varied 
between 2013 December (blue/black points) and 2014 April (red/magenta points).
Other archival points are also shown, to illustrate the variability amplitude of the source.
Note that in the {\it Fermi}/LAT and the source is weak, and the spectrum steep.
In fact these powerful and high--$z$ blazars emit most of their luminosity around $\sim$1 MeV,
and are then better found through hard X--ray surveys rather than $>$100 MeV surveys
(of comparable $\nu F_\nu$ sensitivities).
Adapted from \cite{tagliaferri15}.
}
\label{2149}       
\end{figure}

Fig. \ref{2149} shows data from {\it Swift} and {\it NuSTAR}
of the high--redshift blazars PKS 2149--306 ($z=2.345$) \cite{tagliaferri15}.
Together, {\it Swift}/XRT and {\it NuSTAR} cover the 0.3--70 keV band.
The bottom panel of Fig. \ref{2149} shows the X--ray SED, with the two observations
of {\it Swift+NuSTAR} together with other archival observations.
It can be seen that {\it Swift} is crucial to describe the behaviour of the X--ray spectrum,
that does not change at low energies, while it becomes harder when brighter above
$\sim 4$ keV (13 keV rest frame).
The addition of the {\it Fermi}/LAT makes clear that the hard X--ray behaviour
corresponds to a shift in the high energy peak frequency, that becomes
smaller in the (slightly) lower state. 
In this case we have a behavior opposite to the blazar sequence
(see Fig. \ref{gfos}).

\section{Time domain}

The accumulation of data during the life of {\it Swift} implies that 
a source has the chance to be observed several times, 
with all the three {\it Swift} instruments. 
The most famous blazars (PKS 2155--304, Mkn 421, Mkn 510, 3C 454.3) have 
been observed {\it hundreds} of times.
All data are public, and the {\it Swift} archive is a resource 
for the years to come, still to be fully exploited.
As in many other branches of science,
the amount of data is becoming too large to be analysed by humans
(or, at least, by single humans), and as automatic tools have been
developed for Gamma Ray Bursts, there has been the initiative
of the ASI Science Data Center (ASDC) to offer tools to build the
SED of all sources (not only blazars) with the option to select 
slices of time, or of frequencies.
This is a very important service, publicly available, that I wish to thank.

\section{Accretion disks in blazars}

The non--thermal continuum of blazars often dominates
in the optical--UV bands, making the accretion disk invisible.
And yet, in FSRQs, we do see broad emission lines, that should be
produced by clouds photo--ionized by the disk flux.
The old idea of the beamed continuum being even stronger in BL Lacs, 
such to hide also the emission lines (besides the disk flux) 
is in general wrong, but it may still be true in a few cases.
Now we believe that the ``genuine" BL Lacs intrinsically
lack the broad emission lines, because their accretion disks
are in the low radiative regimes (ion supported tori \cite{rees82}, 
ADAF \cite{narayan97}, CDAF \cite{narayan00}, ADIOS \cite{blandford99}) and
therefore have mass accretion rates below a critical value
in units of Eddington, corresponding to disk luminosities
$L_{\rm disk}/L_{\rm Edd} \lsim 10^{-2}$ \cite{narayan95}, \cite{sbarrato14}.
There is then a ``divide" in terms of $L_{\rm disk}/L_{\rm Edd}$ (or, equivalently, in
terms of the accretion rate $\dot m$ in Eddington units), 
distinguishing BL Lacs and FSRQ \cite{ghisellini09}.
Since we believe that the corresponding parent populations are FR I and FR II radio--galaxies,
they should be characterised by the same divide \cite{ghisellini01}.
All these issues require the knowledge of the mass of the black hole $M_{\rm BH}$.
There are mainly four methods for estimating it.

The first is the popular virial method 
\cite{wandel97}, \cite{peterson14}, requiring the measurement of the FWHM of 
the H$\alpha$ or H$\beta$ or MgII or CIV broad lines, and the
luminosity of the continuum close to the line frequency,
{\it if this is not contaminated by the beamed radiation from the jet}.

The second methods uses the
tight correlation between $M_{\rm BH}$ and the 
velocity dispersion $\sigma$ of the galaxy's bulge or spheroid
\cite{ferrarese00}, \cite{gebhardt00}, \cite{tremain02}.
This is the way in which we can measure $M_{\rm BH}$ of some relatively nearby 
($z<0.4$) BL Lac objects \cite{plotkin11}.

The third method can be applied to relatively nearby BL Lacs whose host galaxy
light can be distinguished and separated from the jet emission \cite{sbarufatti05}.
The black hole mass can then be found through the relation
between the bulge luminosity and $M_{\rm BH}$ \cite{magorrian98}.


The fourth method is the oldest and it is based to the modelling the thermal continuum 
with a disk emission model.
The simplest is a standard, Shakura \& Sunyaev \cite{ss73} model, for which the disk
emits black--body radiation with increasing temperature for smaller radii
\cite{shields78}, \cite{malkan82}, \cite{malkan83}, \cite{sun89}, \cite{zheng95}.
The revival of this method \cite{calderone13} is due to the realisation
that the luminosity of the broad emission lines is a good proxy for the
disk luminosity, that in turn is proportional to the mass accretion rate
$\dot M$ (through the efficiency $\eta$ defined by $L_{\rm disk} =\eta \dot M c^2$).
Therefore, in principle, one can find $M_{\rm BH}$
{\it even when the
peak of the disc emission is not visible, and also when the continuum is
partially contaminated by the jet emission.
}
{\it Swift}/UVOT, with its optical--UV coverage, can be used together with {\it WISE}
or other infrared facilities to study the disk emission at its peak
(or close to it). 
When the peak is visible the statistical error on the derived black hole mass is small
(i.e. less than 50\%), 
in comparison to the virial method.

\begin{figure}
\vskip -1.2 cm
\hskip -0.4 cm
\includegraphics[width=8.4cm,clip]{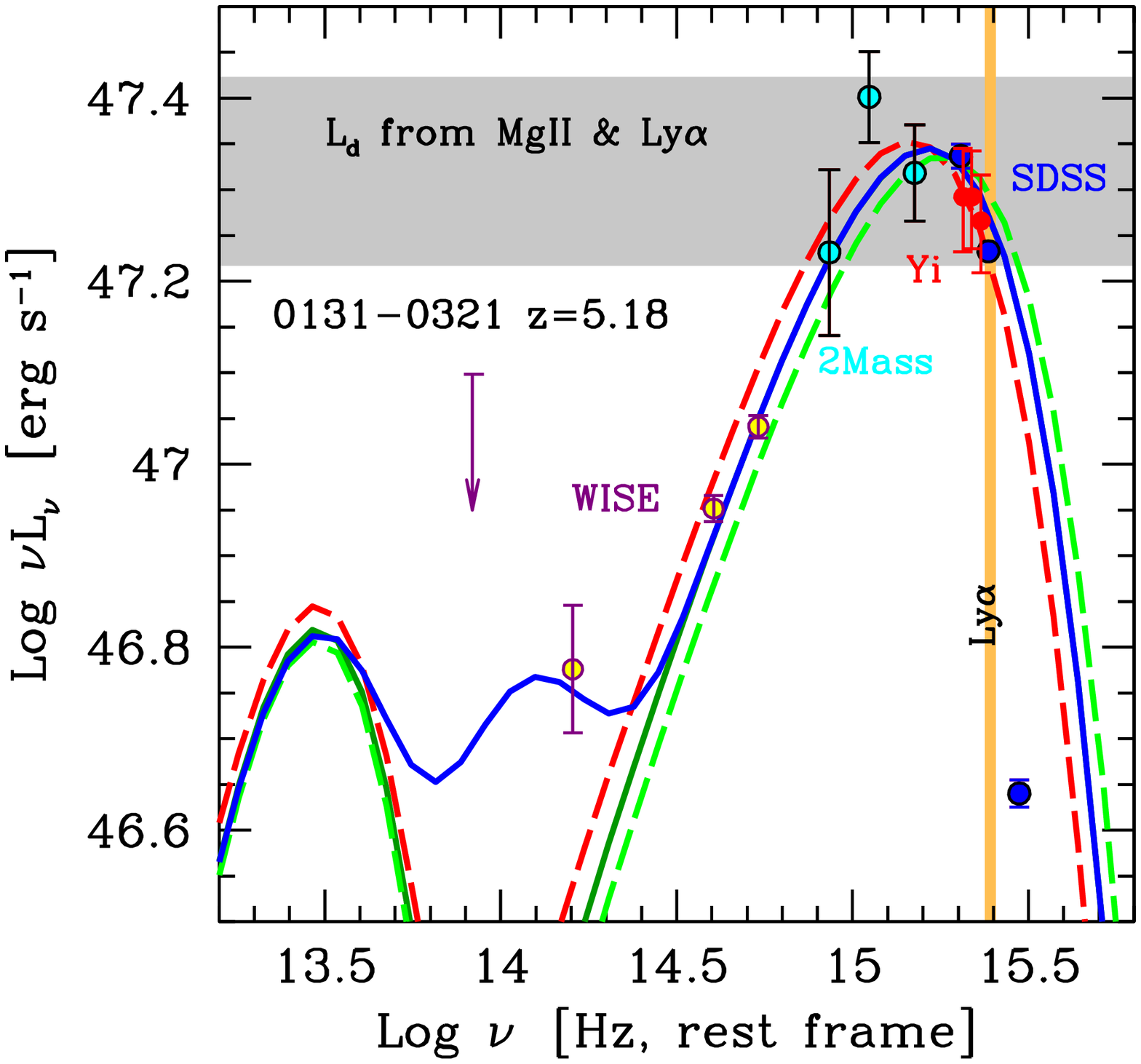}\\
\vskip -0.8 cm
\hskip -0.4 cm
\includegraphics[width=8.4cm,clip]{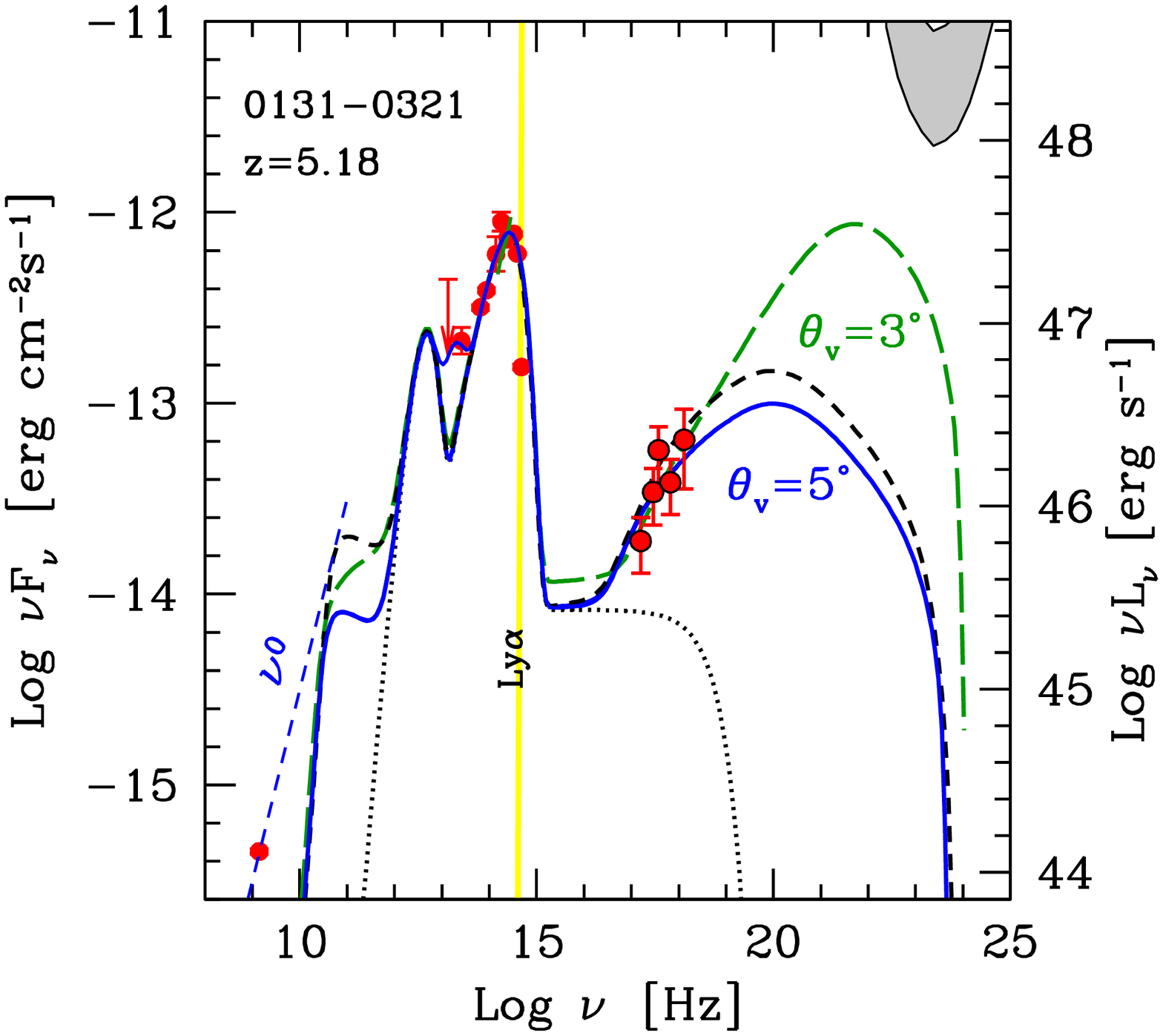}
\vskip -0.5 cm
\caption{
{\bf Top panel:} the IR--optical data (in the rest frame) of the radio--loud quasar SDSS 0131--0321 
($z=5.18$), can be well fitted 
by a standard accretion disk model \cite{ss73}, plus some contribution
from a reprocessing infrared torus \cite{calderone12}. 
The grey stripe indicates the range of possible $\nu L_\nu$ peak luminosities
of the accretion disk, as derived by the broad line luminosities.
The black hole mass is $(11\pm 2)\times 10^9 M_\odot$.
The small uncertainty is due to the well visible (and constrained by the data) peak
of the accretion disk component. 
{\bf Bottom panel:}
entire SED of SDSS 0131--0321 with the recent {\it Swift}/XRT data. 
The SED can be fitted with a leptonic one--zone model, whose jet
is observed with a viewing angle between 3 and 5 degrees.
Since the viewing angle is small, and the jet emission is beamed, there must
exist many other (hundreds) of similar quasars (with the same black hole mass) with jets
pointing in other directions From \cite{ghisellini15}.  
}
\label{0131}       
\end{figure}

The top panel of Fig. \ref{0131} shows an application of this method to the radio--loud source SDSS 0131--0321,
at $z=5.18$ \cite{ghisellini15}.
It shows the IR--UV SED in the rest frame  
and the ocher vertical line indicates the frequency of the hydrogen Ly$\alpha$ line.
The grey stripe is the estimate of the disk luminosity derived by the broad MgII and Ly$\alpha$ lines,
taking into account that we observe only the non--absorbed part of the latter.
The points are archival data. Consider that one of the 2MASS data points includes
the MgII line contribution (and correspondingly the point is above the fitting model).

The solid and dashed lines are three standard disk models, 
with the same $L_{\rm disk}$ and slightly different $M_{\rm BH}$:
9, 11 and 14 {\it billions} of solar masses.
The solid blue line shows also the contribution of two two black--bodies
at different temperatures but similar luminosities, thought to be produced by the torus
(as found in other, radio--quiet sources observed by {\it WISE} \cite{calderone12}).

\begin{figure}
\vskip -0.5 cm
\hskip -0.3 cm
\includegraphics[width=8.6cm,clip]{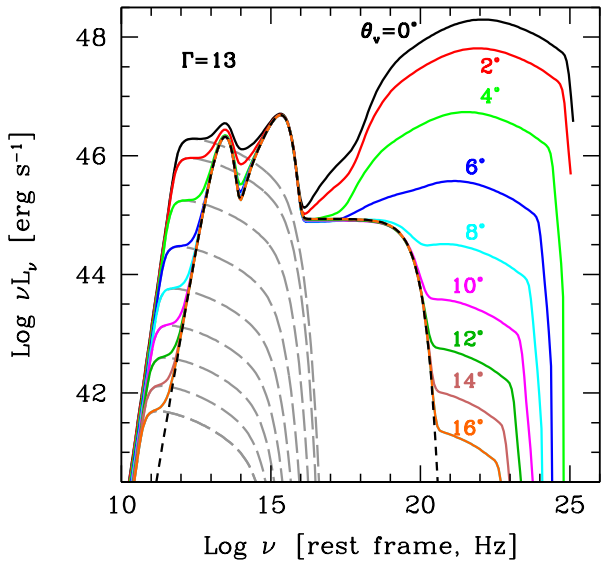}\\
\vskip -1 cm
\caption{A sequence of models illustrating the different
beaming pattern for the synchrotron and the External Compton (EC)
emission that occurs in a powerful blazars, following \cite{dermer95}. 
The very same source is observed at different viewing angles $\theta_{\rm v}$,
as labelled. 
It can be seen that the EC luminosity has a stronger dependence on $\theta_{\rm v}$
than the synchrotron one.
The dashed grey lines are the synchrotron component.
For the chosen $\Gamma$, the proposed divide between blazars and parent 
population occurs at $\theta_{\rm v} =\arcsin(1/\Gamma)=4.4^\circ$.
}
\label{dermer}       
\end{figure}

\section{Confirming the ``blazarness" of high--$z$ blazar candidates}

The simplest way to select sources with jets pointing
close to our line of sight is to consider sources with a flat radio spectrum 
having high values of  {\it radio--loudness} 
$R_{\rm L}=F_{\rm 5\, GHz}/F_{\rm 2500\AA}$, where $F_{\rm 5 \, GHz}$ and $F_{\rm 2500\AA}$ 
are the monochromatic fluxes at 5 GHz and at 2500 $\AA$, respectively.
The rationale of this method is the fact that the flat spectrum radio flux is 
produced by the jet and is amplified by beaming, which strongly depends on the viewing angle,
while the UV flux is not beamed.
However, there are a few caveats:
i) it may happen that {\it both} the radio and the UV are beamed, giving moderate or 
small values of $R_{\rm L}$;
ii) the UV can be thermal emission, and therefore quasi--isotropic, but 
party absorbed by  dust, giving large $R_{\rm L}$ also when the viewing angle is
moderately large.

Therefore we do not have a clear definition of what a blazar is,
beyond the generic requirement that the jet is observed at ``small viewing angles".
But how small?
We proposed that the natural angular scale dividing blazars
from their parent population is the beaming angle, namely $1/\Gamma$.
This definition implies that for each observed blazar there are other $2\Gamma^2$
misaligned sources sharing the same intrinsic properties of the blazars, including the
black hole mass and disk luminosity.
This also implies that we have to find a good method to observationally classify 
blazar candidates as blazars.
Here theory comes to help.
If most of the high energy emission is through inverse Compton scattering 
between relativistic electrons in the jet and seed photons produced outside the jet 
(External Compton, EC), then the radiation pattern is not isotropic even in the comoving frame.
This is because in the comoving frame (by aberration) most external photons are
coming from the forward direction, and head--on scatterings are more 
energetic than tail--on ones.
The observer at a small viewing angles sees not only the usual beaming pattern, but some
extra emission due to the extra power emitted in the forward direction even in the comoving frame.
At large viewing angles, instead, the observer will see less radiation.
In other words, there is a difference between the synchrotron and SSC beamed radiation
(usual beaming pattern) and the EC emission (beaming+some extra pattern), 
making the EC flux more sensitive to the viewing angle \cite{dermer95}.
Fig. \ref{dermer} shows how the model SED changes by changing the viewing angle
$\theta_{\rm v}$.
Notice that from $0^\circ$ to $6^\circ$ the synchrotron luminosity
changes by two orders of magnitude, while the EC one changes by three orders.
Notice also that the X--ray spectrums become steeper for larger viewing angles, 
until the X--ray corona dominates the emission: in the shown example this happens
for $\theta_{\rm v}\gsim 8^\circ$.

This implies that the X--ray flux, relative to the optical and the radio,
becomes a tool to estimate the viewing angle of a candidate blazar,
if the EC component dominates the observed flux.
Fortunately, the EC spectrum is usually much harder than the SSC one,
and we can discriminate.
To summarise: if we want to find new blazars, we first select
flat radio spectra objects with a large radio--loudness (i.e. $>100$),
then we observe it in the X--rays.
If we see a strong X--ray flux relative to the optical and a hard spectrum,
then the object is with good probability a blazar.
The bottom panel of Fig. \ref{0131} shows a borderline case,
because the X--rays, as observed by {\it Swift}/XRT, are not as strong
(relative to the optical) as in the PKS 2149--306 (see the top panel of
Fig. \ref{2149}), but they may have a hard spectrum. 
The source can either be a blazar (corresponding to the predicted spectrum shown by the 
green dashed line) or have $\theta_{\rm v}$ slightly larger than $1/\Gamma$ (solid blue line).
To discriminate we should observe the source at higher X--ray energies, i.e. with {\it NuSTAR}.

From what said it should be clear that this method can work only with FSRQs,
because only in these sources we surely have important sources of external 
photons (broad lines and infrared photons for the torus).
Furthermore, the emission zone should be located within the broad line region,
or at a distance from the black hole smaller than the torus one, for the
EC emission to become effective.

\section{Hard X--ray luminosity function and black hole mass function}

\begin{figure}
\vskip -1.5 cm
\hskip -0.6 cm
\includegraphics[width=8.5cm,clip]{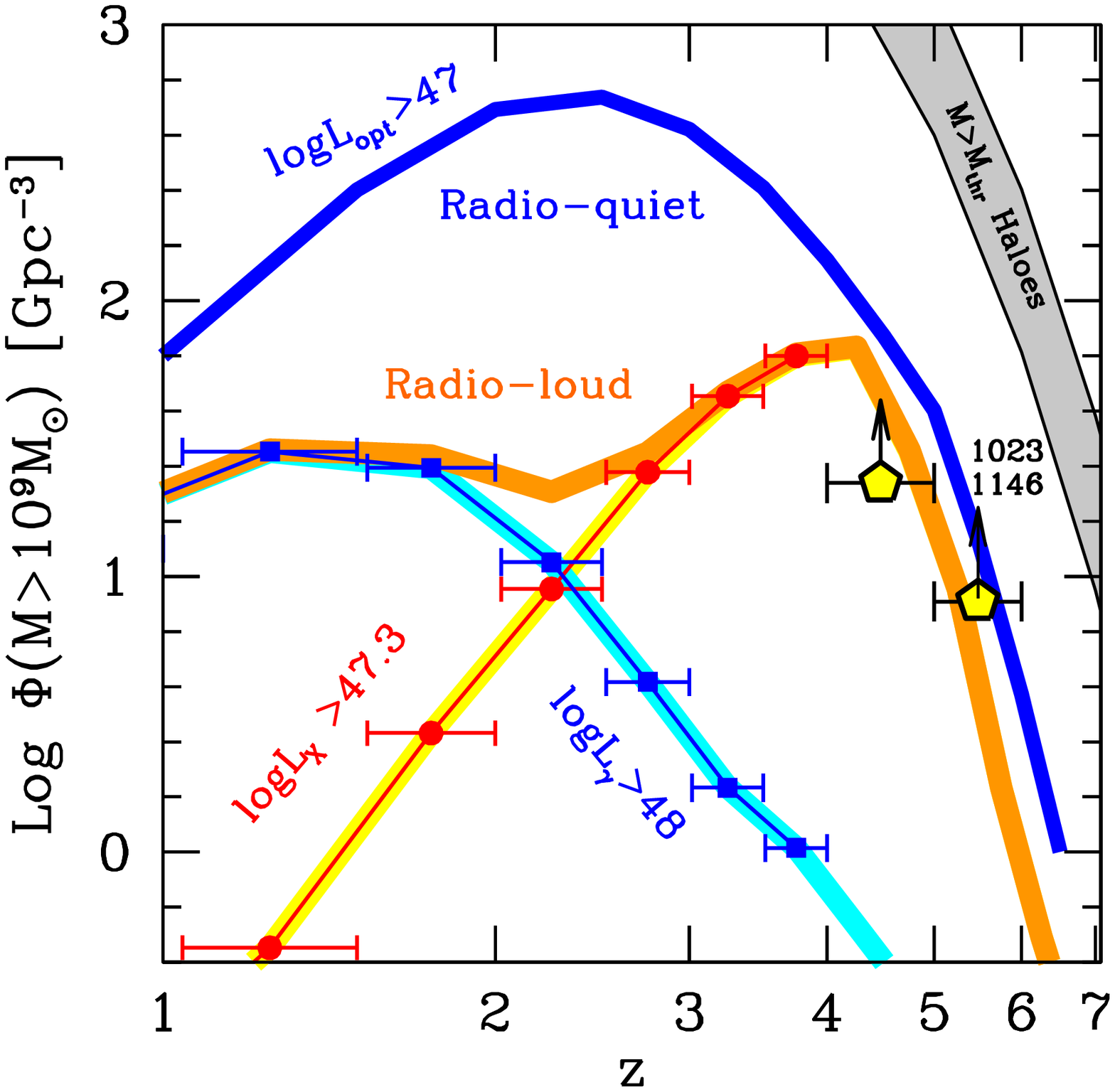}\\
\vskip -1.3 cm
\hskip -0.6 cm
\includegraphics[width=8.5cm,clip]{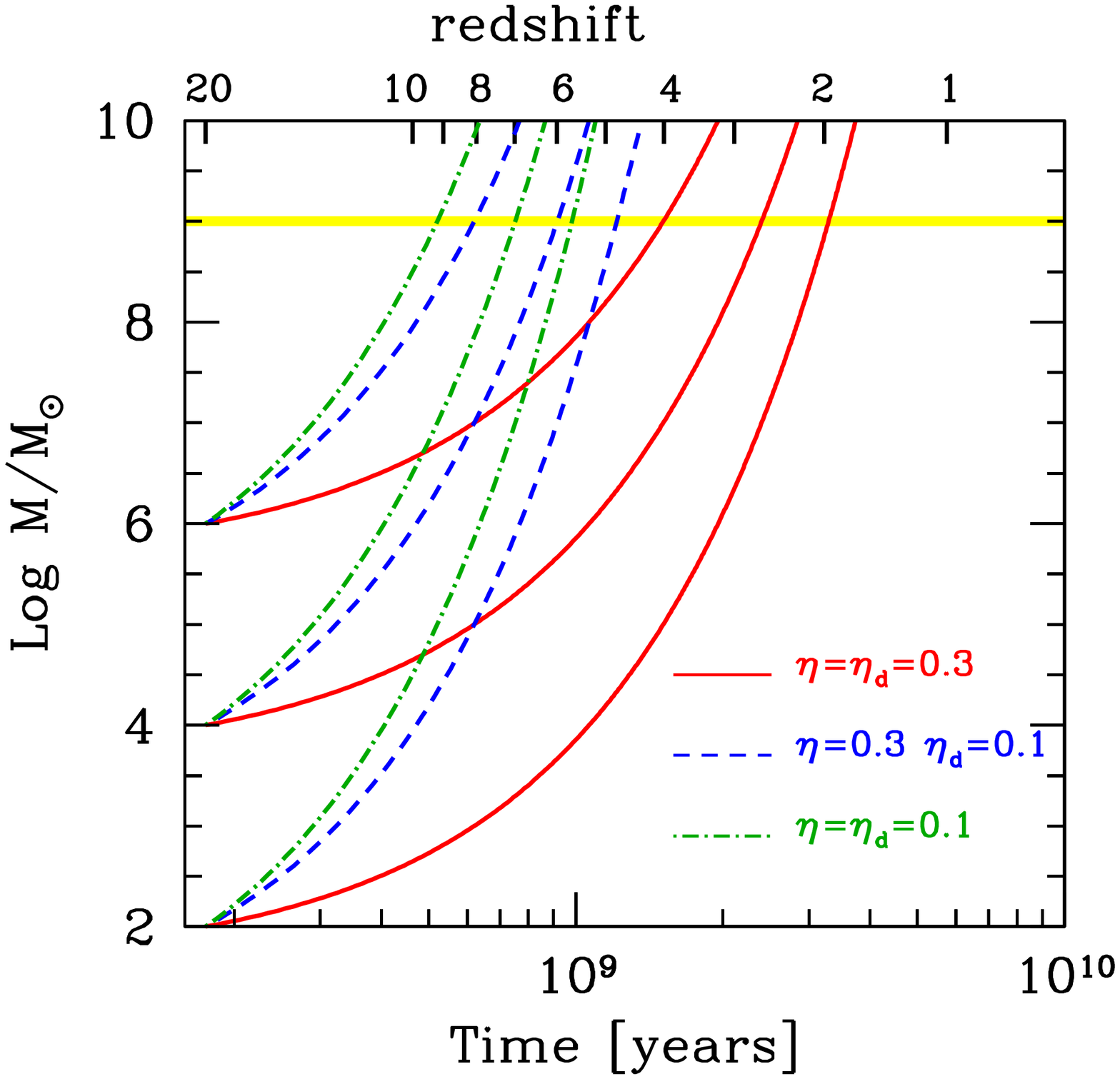}
\vskip -0.6 cm
\caption{ {\bf Top:}
Comoving number density of  blazars powered by ``heavy and active" black holes 
($M>10^9M_\odot$, $L_{\rm disk}/L_{\rm Edd}>0.1$) as a function of redshift.
The orange stripe is derived by integrating the [15--55 keV]
luminosity function (LF) (\cite{ajello09}, 
as modified in \cite{ghisellini10}) 
above $L_{\rm BAT}=2\times 10^{47}$ erg/s, and multiplying
the derived density by $2\Gamma^2=450$ (i.e. $\Gamma=15$).
The light blue stripe is derived by integrating the 
$\gamma$--ray LF \cite{ajello12} above $L_\gamma=10^{48}$ erg/s. 
The blue stripe is derived integrating the LF of
radio--quiet quasars \cite{hopkins07} above $L_{\rm opt}=10^{47}$ erg/s,
as labelled (see also \cite{willott00} and \cite{volonteri11}).
The grey stripe, described in \cite{ghisellini10}
can be considered as an upper limit. 
The (yellow) pentagon labelled 1023 1146 is the density inferred from the existence of two blazars,
B2 1023+25 \cite{sbarrato13}, and SDSS J1146+4037 \cite{gg14b}
at $z>5$ in the region of the sky covered by the SDSS+FIRST surveys.
Adapted from \cite{sbarrato15}.
{\bf  Bottom:}
$M_{\rm BH}$
as a function of time (bottom axis) and redshift (top axis). 
Accretion starts at $z=20$ onto a black hole seed of $10^2M_\odot$,
 $10^4M_\odot$ or $10^6M_\odot$, with different efficiencies, as labeled.
The larger $\eta_{\rm d}$, the smaller the amount of accreted mass
needed to produce a given luminosity, and the longer the black hole 
growing time. If part of the accretion energy goes into launching a jet,
however, $\eta_{\rm d}<\eta$ and the growth time decreases. From \cite{gg13b}.
}
\vskip -0.7 cm
\label{fi}       
\end{figure}

Ajello et al. \cite{ajello09} studied all blazars detected by {\it Swift}/BAT after the first 
3 years of operations (see also \cite{baumgartner13}, \cite{cusumano10} for more recent 
BAT catalogs), and constructed the corresponding blazar luminosity function.
The number of detected blazars was very limited: 38 in total: 26 FSRQ and 12 BL Lacs
(excluding the Galactic plane from the analysis).
Of these, 10 blazars (all FSRQs) are at $z>2$, and 5 at $3<z<4$.
All these 10 blazars have a 15--55 keV luminosity  
$L_{\rm BAT}>10^{47.2}$ erg s$^{-1}$ \cite{ajello09}.
We \cite{ghisellini10} studied these 10 high--$z$ blazars and found that the black hole mass
of all of them exceeded $10^9 M_\odot$.
Therefore the number density of blazars at $z>2$ with $L_{\rm BAT}>10^{47.2}$ erg s$^{-1}$
can be used to estimate the number density of heavy black holes at these redshifts.
Using the average value of $\Gamma$ we can reconstruct the total comoving number density of jetted sources
having $M_{\rm BH}>10^9 M_\odot$.
The result shown in Fig. \ref{fi} is intriguing.
The density of active and heavy black holes in jetted sources, as derived by using BAT data,
peaks at $z\sim 4$.
We stress that this is the density of {\it active} black holes, namely having $L_{\rm disk}>0.1 L_{\rm Edd}$.
We can contrast this density profile with the one derived by the blazars detected by {\it Fermi}/LAT
and studied in \cite{ajello12}, shown in Fig. \ref{fi} as the light blue line.
In this case the mass density profile of heavy and active black holes peaks at $z \sim 1.5$,
but the peak is at a lower density value.
Since we are considering strongly accreting sources, we expect that most of their 
jet luminosity is emitted in the hard X--rays or in the $\gamma$--ray bands,
therefore in the {\it Swift}/BAT or in the {\it Fermi}/LAT bands.
We conclude that most heavy black holes in jetted sources are indeed formed at $z\sim 4$.

By integrating the optical luminosity function of \cite{hopkins07} above $L_{\rm opt}=10^{47}$ erg s$^{-1}$,
we can have an estimate of the density of heavy and active black holes for radio--quiet quasars
(the luminosity limit corresponds to nearly the Eddington luminosity for a $10^9 M_\odot$ black hole),
shown by the blue line in Fig. \ref{fi}.
This profile peaks at $z\sim 2.5$
(see Fig. 2 in \cite{ghisellini11} for different $L_{\rm opt}$ thresholds).
The conclusion is that very heavy black holes in jetted sources form earlier than
very heavy black holes in radio--quiet sources.
Is this because jets are born preferentially in heavy black hole systems,
or, on the contrary, it is the jet that helps a fast growth of the black hole?

As Jolley \& Kuncic \cite{jolley08}  suggested, the available gravitational energy
of the infalling matter could be used not only to heat the disk, but also to
amplify the magnetic field necessary to launch the jet.
In this case the disk is colder, and becomes Eddington limited for a higher accretion rate.
If there is a large reservoir of matter that can be accreted, we can have large $\dot M$
and relatively small $L_{\rm disk}$ even if the total efficiency of the accretion is $\eta=0.3$.
In practice, the total efficiency $\eta=\eta_{\rm d}+\eta_{\rm jet}$ 
is the sum of the radiative efficiency $\eta_{\rm d}$ and the jet efficiency $\eta_{\rm jet}$.

The bottom panel of Fig. \ref{fi} shows the black hole mass as a function of time (or redshift,
upper x--axis) for accretion limited by the Eddington rate and for 
$10^2$, $10^4$ and $10^6 M_\odot$ of the seed black hole, assumed to be at $z=20$
(the figure is taken from \cite{sbarrato15}).
For the lowest value of the seed, and for the usual $\eta=\eta_{\rm d}=0.1$ value,
it is not possible to have $10^9 M_\odot$ black holes at $z\gsim 2$.
Instead we do find larger masses at higher redshifts.
Increasing the seed black hole mass does not help much: even
a $10^6 M_\odot$ seed reaches $10^9 M_\odot$ not earlier than $z=4$.
We must assume a lower efficiency $\eta$.
But this contrasts with the general idea that jets are associated with 
black holes spinning rapidly, close to the maximum value.
In this case the innermost stable orbit approaches the gravitational radius $R_{\rm g} =GM_{\rm BH}/c^2$,
and the efficiency is close to the maximum value (which is $\eta=0.42$ in principle, but 
only $\eta=0.3$ in reality, see \cite{thorne74}).
The problem can be solved if we assume that even in $\eta=0.3$ systems, most of the gravitational
energy is used to amplify the magnetic field necessary to tap the rotational energy of the hole.
The ``disk efficiency" $\eta_{\rm d}$ can then be smaller than 0.1: to produce a given disk luminosity,
the mass accretion rate is correspondingly larger, and the black hole can then grow faster
(blue dashed lines in Fig. \ref{fi}).

But why the number density of heavy black holes derived from the $\gamma$--ray luminosity
function peaks at a much smaller redshift, similar to the peak of the radio--quiet ones?
Are these black holes in their growth age, or are they resurrected black holes,
e.g. by a merging event? 
We do not know yet.

In any case, there seems to be {\it two epochs} of formation of heavy black holes.
One, at $z\sim 4$, in which the majority of $M_{\rm BH}>10^9 M_\odot$ 
with jets is born and grows, the other, at $z\sim 2$ where the majority
of heavy black holes in radio--quiet quasars are born.
Since we are concerned only with heavy black holes, that are a minority of the total, 
this might not affect the radio--loud fraction of the entire population of quasars,
that is seen to decrease with redshift (but not with luminosity) \cite{jiang07}.

These results are a direct consequence of the {\it Swift}/BAT survey. 
The fact that it includes blazars at higher redshifts than {\it Fermi}/LAT is
the consequence of two effects: i) increasing the redshift, the Compton peak of the SED moves
closer to the observed high energy X--ray band, and ii)
more powerful sources have the Compton peak located (in the rest frame) at 
somewhat smaller energies (i.e. close to 1 MeV), in general agreement with the blazar sequence
(see Fig. 11 of \cite{ghisellini10}).

\begin{figure}
\includegraphics[width=8.5cm,clip]{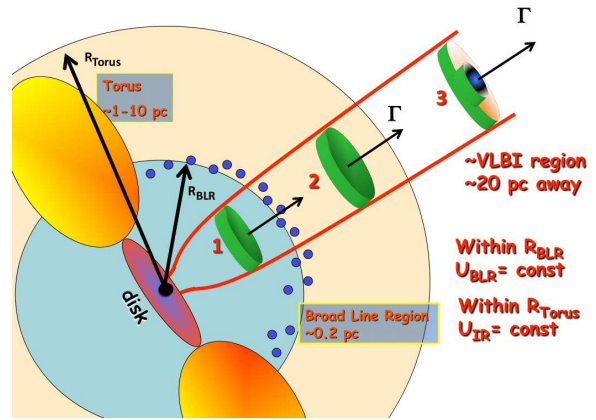} \\
\caption{
Cartoon of the possible location of the emitting region.
The most economic location is within the BLR (1): 
there the inverse Compton process can benefit of the 
photons produced by the BLR.
If the location of the emitting region is beyond the BLR (2),
but within $R_{\rm torus}$, there still is the benefit
of using the infrared photons produced by the torus.
Beyond $R_{\rm torus}$ (3) the energy density of the external radiation drops 
rapidly, and the main source of photons for scattering 
is the internally produced synchrotron radiation.
The emitting region, in this case, must be a small part of the
jet, otherwise the variability timescales become too long.
}
\label{cartoon}       
\end{figure}

\section{Location of the emitting region}

There is an ongoing important debate about the location of the emitting region,
and if the one--zone model is an over--simplification or instead if it is a good
representation of reality.

First, we have to be aware that the jet has surely several
emitting regions: we do see some of them in the VLBI observations, and
it is very likely that this ``knotty" structure exists also at the smallest scale.
On the other hand, if the jet is accelerating, the beaming angle of the most inner
parts is very large, implying that the corresponding luminosity is diluted in a large
solid angle, and therefore not visible in blazars (but it could be in radio--galaxies,
\cite{ghisellini12}).
However, there could be several emitting zones also when $\Gamma$ has reached its
final value.
After all, this is what predicted by the ``internal shock model" for blazars
\cite{rees78}, \cite{spada01}.
And yet we often see coordinated variability at different frequencies,
implying that there is one region that dominates the bolometric luminosity.
{\it Swift} gave a fundamental contribution to this issue.

{\bf Within the BLR ---}
The variability timescales $t_{\rm var}$ are often short, of the order of few hours
\cite{tavecchio10}, and imply rather small dimensions of the emitting region,
$R \lsim ct_{\rm var} \delta/(1+z) \approx 4\times 10^{15}(t_{\rm var}/3\, {\rm h})(\delta/10)/(1+z)$ cm
($\delta$ is the relativistic Doppler factor). 
In turn, for conical jets of aperture angle $\psi\sim 0.1$ rad, we derive that the 
emitting region is at $R_{\rm diss}\sim 4\times 16$ cm  from the black hole. 
Since the broad line region is supposed to be at a distance 
$R_{\rm BLR} \sim 10^{17} (L_{\rm disk}/10^{45} \,{\rm erg/s})^{1/2}$ cm,
most FSRQs should have their emitting region preferentially
within the BLR (zone 1 in Fig. \ref{cartoon}).
In  this case there should be a well defined signature in the $\gamma$--ray
spectrum of blazars, since the photons of the broad lines at high frequency (e.g. HeII)
could be the targets for the $\gamma$--$\gamma \to$ e$^\pm$ process.
This should produce an absorption feature in the high energy (GeV) spectrum 
\cite{poutanen10}, \cite{stern11}.

{\bf Beyond the BLR but within the torus ---}
Another possible location for the most active part of the jet of FSRQs
is outside the BLR, but within $R_{\rm torus}$ \cite{blazejowski00}.
By ``torus" we mean a reprocessing region similar to the one in radio--quiet
quasars, intercepting 10--40\% of $L_{\rm disk}$ and re--emitting it in the IR,
probably with a multi temperature blackbody spectrum, the hottest being $T\sim 2000 K$,
the sublimation temperature of the dust.
This is the ``2" zone in Fig. \ref{cartoon}. 
Both $R_{\rm BLR}$ and $R_{\rm torus}$ should scale as $L^{-1/2}_{\rm disk}$.
This implies that within $R_{\rm BLR}$ the radiation energy density $U_{\rm BLR}$ is constant.  
Also between $R_{\rm BLR}$ and $R_{\rm torus}$ the radiation energy density $U_{\rm torus}$
is constant  but with a reduced value: $U_{\rm BLR}>U_{\rm torus}$.
Since it is likely that the magnetic field $B$ scales as $R^{-1}$, corresponding
to a constant Poynting flux ($\propto R^2 \Gamma^2 B^2$),
an emitting zone further out in the jet implies a less magnetised zone.
A given Compton to synchrotron luminosity ratio can usually be produced 
in two well defined zones, one within $R_{\rm BLR}$ and one between $R_{\rm BLR}$ and $R_{\rm torus}$
\cite{gg09}.
To choose, apart from the expected variability timescale, we can use the peak frequency of
the Compton hump: if smaller than a few MeV it may indicate the need for IR seed photons.

{\bf Beyond the BLR and the torus ---}
In some blazars there seems to be a correlation between $\gamma$--ray flares,
radio flares, and  a switch in the polarization angle \cite{jorstadt13}, \cite{marscher08}.
This suggests that the emitting region is located parsecs away from the black hole,
in the VLBI region. 
This contrasts with the short $t_{\rm var}$, and also with the cooling timescale, 
that should be long at that distances (small magnetic field, no important sources of external photons).
And yet, the ultrafast variability seen in the TeV band is a puzzle.
Two TeV blazars, Mkn 501 and PKS 2155--304 showed  significant (factor 2) 
variations of their TeV flux in $t_{\rm var}=$3--5 minutes \cite{albert07}, \cite{aharonian07}.
This is already difficult to explain \cite{begelman08}, 
\cite{giannios09}, \cite{gg09c}, \cite{marscher10}, 
but the real puzzle came when PKS 1222+216, at $z=0.431$,
was also observed to vary, at a few hundreds GeV, in $t_{\rm var}\sim 10$ minutes \cite{aleksic11b},
corresponding to a size $R<c t_{\rm var}\, \delta/(1+z) \simeq 5\times 10^{14} (\delta/20)$ cm.
PKS 1222+216 is a FSRQ with broad emission lines, and this extremely small TeV emitting region 
cannot be located within the BLR, whose photons would absorb the emission above $\sim$20 GeV.
It must be located outside.
This remains true even if the BLR has a flattened geometry (as suggested by e.g. \cite{shields78},
\cite{jarvis06} and \cite{decarli11}). 
The TeV emitting region of PKS 1222+216 can be located between $R_{\rm BLR}$ and $R_{\rm torus}$,
but must be much smaller than the cross sectional radius of the jet at these distances,
or the jet itself must shrink, due to  
strong recollimation and focusing of the flow (e.g. \cite{stawarz06}; 
\cite{bromberg09}, \cite{nalewajko09}).
Alternatively we may have complex reconnection events \cite{giannios13};
or even a photon to axion transition, to survive the $\gamma$--$\gamma$ process \cite{tavecchio12}.

\section{Jets and accretion}

In their pioneering work \cite{rawlings91} Rawlings \& Saunders found
that the radio--lobes of AGNs, to exist, require an average power 
that is of the same order of the accretion disk luminosity.
Uncertainties were large, since there was no idea of the
contribution of the protons to the total lobe energetics.
Attempts to measure the jet power continued, modelling
the SED using the VLBI radio size and a limit on the $\delta$--factor 
from the requirement not to exceed, by the SSC process, the
observed X--ray flux \cite{celotti93}, \cite{celotti97}.

In a similar way, modelling the radio--optical and X-ray emission
at relatively large jet scale, as resolved by {\it Chandra},
allowed to estimate the jet power at these scales \cite{ghisellini01b}, and to compare
it with the values found for the compact jet \cite{tavecchio07}.

Another more recent way to indirectly measure the jet power is through 
X--ray cavities seen in (relatively nearby) radio--galaxies,
coinciding with their radio--lobes: 
The mechanical power $P_{\rm cav}$ of the cavity can be derived by setting  
$P_{\rm cav}=PV/t_{\rm age}$, where $t_{\rm age}$ is the rise--time of the cavity,
$P$ is the pressure and $V$ is the volume of the cavity \cite{fabian12}.

With the advent of {\it CGRO} and now {\it Fermi} we at last knew that 
most of the power of blazars was emitted at high energies.
In powerful FSRQs, the inverse Compton luminosity can be larger than the synchrotron one
by up to 3 orders of magnitude.
We can then derive the power spent by the jet to produce the 
observed jet bolometric luminosity $L_{\rm jet}$ (corresponding
to $L^\prime_{\rm jet}$ in the comoving frame).
If the processes are synchrotron and SSC (isotropic in the comoving frame) we have,
{\it for one jet}:
\begin{equation}
P_{\rm rad} =  {L^\prime_{\rm jet} \over 4\pi } \, \int \delta^4 d\Omega =  
{4\over 3}\Gamma^2 L^\prime_{\rm jet}  =
{4\over 3}{L_{\rm jet}\over \Gamma^2}  
\end{equation} 
where the last equality is for viewing angles $\theta_{\rm v}\sim 1/\Gamma$,
for which $\delta=\Gamma$.
This is the entire power carried by the produced radiation, at all angles.
If most of the luminosity is produced by the external Compton process, the beaming
pattern is not $\propto \delta^4$, but $\propto \delta^4 (\delta/\Gamma)^2$
\cite{dermer95}, \cite{georganopoulos01}.
Setting $\langle L^\prime_{\rm jet} \rangle$ as the angle--averaged luminosity in the comoving frame,
we have {\it for one jet} \cite{ghisellini10b}:
\begin{eqnarray}
P_{\rm rad} &=& {\langle L^\prime_{\rm jet}\rangle  \over 4 \pi}\,
\int_{4\pi} {\delta^6(\theta) \over \Gamma^2} d\Omega \sim 
{16\over 5} \Gamma^2 \langle L^\prime_{\rm jet} \rangle \nonumber \\
&\approx& 
{16 \, \Gamma^4 L_{\rm jet} \over 5 \, \delta^6(\theta_{\rm v}) } \sim 
{16 \, L_{\rm jet} \over 5 \, \Gamma^2 }
\end{eqnarray}
Again, the last equality is valid for  $\theta_{\rm v}\sim 1/\Gamma$.
We see that the difference between the two cases, for blazars (i.e. when $\delta\sim \Gamma$), 
is only in the numerical coefficient.

To estimate the power $P_{\rm rad}$ we need to know the bolometric jet luminosity and
the value of $\Gamma$.
The latter can be derived by modelling, or by assuming that the values derived from the 
apparent superluminal velocity occurring at the VLBI scale (i.e. 10 pc)
are the same of the emitting region.
It is therefore a very robust quantity, almost model--independent.
On the other hand, $P_{\rm rad}$ is only a lower limit to the real jet power:
if all of the jet power is spent to produce the radiation we see, 
then the jet should stop, and could not produce the superluminal
blobs or energise the extended radio structure.
In low powerful, TeV BL Lacs, something of this kind may indeed happen \cite{ghisellini05}:
these are sources requiring the highest $\Gamma$ in the emitting
region, and yet they do not show superluminal motion, nor strong extended
structure.

We want to compare $P_{\rm rad}$ with the disk luminosity $L_{\rm disk}$.
We then need a sample of blazars detected in the $\gamma$--ray band (where
most of $L_{\rm jet}$ is emitted) and for which we can reliably estimate $L_{\rm disk}$.
This sample has been assembled rather recently by Shaw et al. \cite{shaw12}, \cite{shaw13},
that observed spectroscopically hundreds of blazars (both FSRQs and BL Lacs) 
detected by {\it Fermi}/LAT.

\begin{figure}
\vskip -0.5 cm
\hskip -0.2 cm
\includegraphics[width=8.5cm,clip]{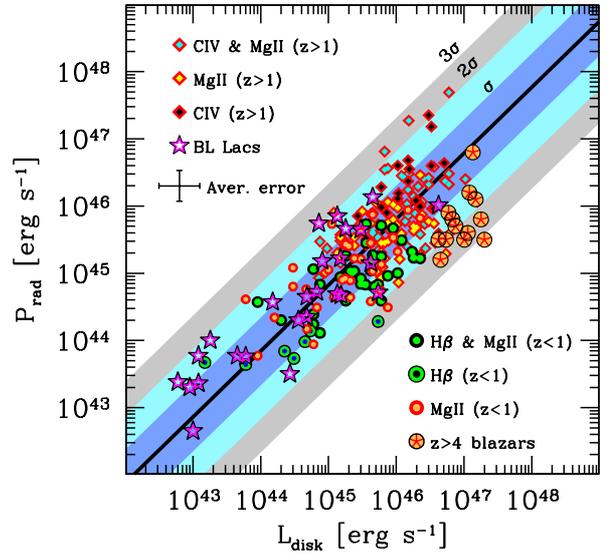} \\
\vskip -0.9 cm
\caption{
The power spent by the blazar jet to produce the radiation we see,
$P_{\rm rad}$ is comparable with the accretion disk luminosity $L_{\rm disk}$.
All shown blazars have broad lines, used as a proxy for the disk luminosity, and
are detected by {\it Fermi}/LAT, and thus have a well determined bolometric 
non--thermal luminosity $L^{\rm obs}_{\rm bol}$, from which 
$P_{\rm rad}\sim L_{\rm jet}/\Gamma^2$ is derived.
Therefore the result $P_{\rm rad}\sim L_{\rm disk}$ is quasi model--independent,
since $\Gamma$ is rather well determined.
The plotted $P_{\rm rad}$ refer to {\it two} jets.
Adapted from \cite{gg14}, with the addition of all known $z>4$ blazars.
}
\label{prld}       
\end{figure}

We \cite{gg14}, \cite{gg15}
have selected all the 217 objects showing at least one broad emission line
in their spectrum.
Among those we have 26 blazars classified as BL Lacs according to the classical 
definition (i.e. equivalent width smaller than 5 $\AA$)
but that indeed have a broad emission line in their spectrum.
Therefore they belong to the low line luminosity tail of FSRQs rather than being
classical BL Lacs.
We constructed the overall SEDs of all objects, calculated the bolometric
jet luminosity
$L_{\rm jet}$, and applied a one--zone model to fit the data from the mm to the 
high energy $\gamma$--rays.
We could then derive the bulk Lorentz factor $\Gamma$ for each source,
finding a narrow distribution ($10<\Gamma<17$).
This agrees with the values found through superluminal motion.
Therefore we could reliably estimate $P_{\rm rad}$ for all sources:
at this stage, the applied model is used only for estimating $\Gamma$.

The disk luminosity could be estimated through the observed emission lines, using the
average templates of \cite{francis91} or \cite{vandenberk01} and 
then multiplying by the average ratio $L_{\rm disk}/L_{\rm BLR}\sim 10$,
or using the value directly listed in \cite{shaw12}, or using
the value found fitting the IR--UV SED with a Shakura \& Sunyaev disk spectrum.
We checked that the three methods agreed. 
In the few cases for which they did not, we choose the disk--fitting value.
Fig. \ref{prld} shows the result: $P_{\rm rad}\sim L_{\rm disk}$ 
with a probability $P < 10^{-8}$ to be random, 
even taking into account the common redshift dependence.
The coloured stripes indicate 1, 2, and 3$\sigma$ (vertical) dispersion ($\sigma=$0.5 dex).
Just for curiosity, we plot also the values for a sample of $z>4$ FSRQs
having good optical--UV spectral coverage (hence a reliable $L_{\rm disk}$)
and X--ray data, used to estimate $P_{\rm rad}$. 
Since they have not (yet) been detected by {\it Fermi}/LAT, 
their inverse Compton hump luminosity is uncertain, and
their real $L_{\rm jet}$ could be larger.
From the found correlation we can conclude that the lower limit on the jet 
power is of the same order of the disk luminosity, and therefore that 
{\it the real jet power should be larger than $L_{\rm disk}$}.

The problem is to find the real $P_{\rm jet}$.
In the past, there have been four main uncertainties concerning
the estimate of $P_{\rm jet}$ by using emission models to fit the spectra:
\begin{enumerate}

\item {\it The emission model could be hadronic, rather that leptonic.}

In this case one needs very energetic protons, that pair--produce 
in photo--meson interactions. The jet power, as calculated in \cite{bottcher13}
{\it is greater} than in leptonic models.

\item {\it The emitting region could be far away, with SSC as the most important
emitting process.}

Usually, a simple SSC does not fit well the SED of FSRQs.
Besides, if the emitting region is beyond $R_{\rm torus}$, there are no important sources
of external photons, and presumably the magnetic field is smaller. 
This means that the synchrotron and inverse Compton scattering process become 
less efficient radiators. 
To produce the radiation we see, we need more particles, and the jet kinetic power 
is larger.

\item {\it The assumption of one proton per electron makes the
proton kinetic energy dominating $P_{\rm jet}$.
But most of the electrons, for energy distribution $N(\gamma)\propto \gamma^{-p}$
with $p>1$, lies at the low energy end. If there is low energy
cut--off, we might not notice it in the data.}

In FSRQs, the EC process dominates, making the soft X--ray spectrum very hard.
This part of the SED is produced by {\it low energy} electrons scattering either 
broad line photons, mainly of the hydrogen Ly$\alpha$ line, or IR photons
from the torus.
We do have control of the low energy tail of the particle distribution.

\item {\it The assumption of one proton per electron could be wrong 
because of electron--positron pairs.}

Pairs cannot be produced (in appreciable number) in the emitting region, otherwise they would
reprocess the spectrum, especially at the low X--ray frequencies, making the
X--ray spectrum softer than observed.
There is the possibility to produce pairs at the base of the jet, in a region with
small $\Gamma$, whose radiation is overwhelmed by the main emitting region in blazars,
but could be visible in radio galaxies. 
However, the SED is required to be finely tuned, to produced a sufficient number of
pairs \cite{ghisellini12}, and this is unlikely.
See also \cite{sikora00}, \cite{celotti08}, \cite{ghisellini10b}
for additional arguments concerning the presence of pairs. 

\end{enumerate}
%


In addition to these arguments, and in completely independently way
we also have the result of Nemmen et al. \cite{nemmen12},
that estimated $P_{\rm rad}$ and $P_{\rm jet}$ for blazars and Gamma Ray Bursts,
finding that they lie on the same correlation, indicating $P_{\rm jet}\sim 10 P_{\rm rad}$.
We recover their result if we assume one proton per electron.

\begin{figure}
\vskip -0.2 cm
\hskip -0.5 cm
\includegraphics[width=9cm,clip]{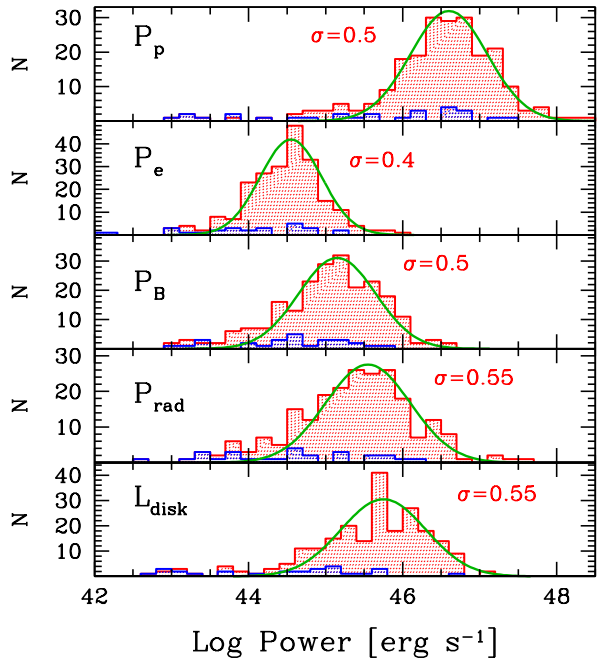}
\vskip -0.2 cm
\caption{Distribution of the different forms of jet powers compared to the  
disk luminosity. The jet powers assumed {\it two} jets. 
The darker (blues) hatched regions correspond to the ``BL Lacs" of the sample.
$P_{\rm p}$ and $P_{\rm e}$ are the kinetic power of emitting electrons and of the (cold) protons,
$P_{\rm B}$ is the Poyinting flux,
$L_{\rm disk}$ is the disk luminosity. From \cite{gg14}
}
\label{powers}       
\end{figure}

Fig. \ref{powers} shows how the different forms of powers are distributed:
$P_{\rm B}$ is the Poynting flux, $P_{\rm e}$ is the kinetic power of the
emitting electrons (thus including their relativistic random energy),
and $P_{\rm p}$ is the kinetic power of protons, assumed to be cold in the comoving frame.
One can see that to produce $P_{\rm rad}$, the Poynting flux and the electron kinetic
power are not sufficient. 
One needs another reservoir of power, and the simple assumption is that this is
provided by protons.

\begin{figure}
\vskip -0.5 cm
\hskip -0.5 cm
\includegraphics[width=8.8cm,clip]{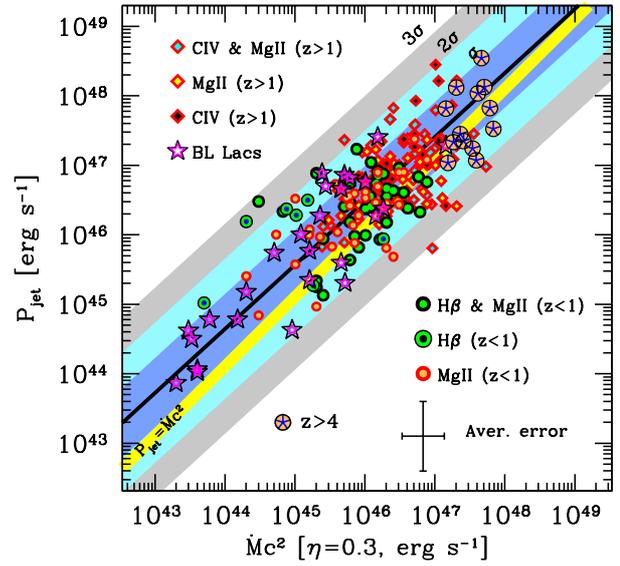}
\vskip -0.8 cm
\caption{
The total jet power $P_{\rm jet}$ as a function of the mass accretion rate, assuming
an accretion efficiency $\eta$ (defined through $L_{\rm disk}=\eta \dot M c^2$) equal to 0.3,
a value appropriate for maximally spinning black holes \cite{thorne74}.
The yellow line is the equality line, while the black line is the best fit.
{\it The jet power is of the same order of, and possibly larger than, $\dot M c^2$.}
Therefore, despite the strong correlation between the jet power and the accretion rate, 
the latter is not enough to power the jet. 
Another source of power is needed, such as the extraction of the rotational energy of the spinning 
black hole.
Adapted from \cite{gg14}, with the addition of all known $z>4$ blazars.
}
\label{pjdotm}       
\end{figure}

Finally, Fig. \ref{pjdotm} shows the total jet power as a function of $\dot M$,
The latter is found assuming that $L_{\rm disk}=\eta\dot M c^2$, with $\eta=0.3$,
i.e. for a maximally efficient accretion and maximally rotating black hole.
The yellow stripe indicates equality, while the black line is the best fit
of the correlation. 
We find that $P_{\rm jet}\gsim \dot M c^2$, and yet it is correlated with it.
This is an apparent paradox: the fact that it correlates let us think
that the jet is powered by accretion, while the fact that its power is
greater than $\dot M c^2$ implies that this is not possible.

The solution of this paradox is the following:
part of the gravitational energy of the infalling matter
is used to amplify seed magnetic fields up to equipartition with the 
mass energy density $\sim \rho c^2$ of the matter accreting 
at the rate $\dot M$. 
We then have $B^2\propto \rho \propto \dot M$.
According to the Blandford \& Znajek process 
\cite{blandford77}, the jet power depends on  $(a M B)^2$, the square of the product 
of the hole spin, mass and the magnetic field at the horizon. 
Therefore $P_{\rm jet} \propto a^2 B^2 \propto  a^2 \rho \propto \dot M$.
This explains why $P_{\rm jet}$ correlates with the accretion power.

To explain why $P_{\rm jet}\gsim \dot M c^2$ we are forced to assume
that the jet production process extracts energy not only from accretion,
but mostly from the rotational energy of the black hole.
This is done by the magnetic field that must be thought as a {\it catalyst} for the process. 
Since $P_{\rm jet}\gsim \dot M c^2 >L_{\rm disk}$, the extraction of 
the rotational energy must be extremely efficient.
This  is fully consistent with  general relativistic magnetohydrodynamic
numerical simulations \cite{tchekhovskoy11},
in which the average outflowing power in jets and winds reaches 140\% of $\dot M c^2$
for dimensionless spin values $a=$0.99.
Occasionally, the magnetic energy density can exceed the energy density $\sim \rho c^2$ of
in the vicinity of the last stable orbit, and the
accretion is temporarily halted {\cite{tchekhovskoy11}, \cite{tchekhovskoy14}, \cite{zamaninasab14},
providing a way to explain the observed variability.








\end{document}